\documentclass[fleqn,10pt]{SelfArx} 
\usepackage[numbers,sort&compress]{natbib}
\usepackage[font={small,singlespacing}]{caption}
\usepackage{setspace}
\usepackage{hyperref}
\usepackage{lettrine}
\usepackage{booktabs} 
\usepackage{array} 

\definecolor{color1}{RGB}{39,45,102} 
\definecolor{color2}{RGB}{240,240,240} 

\newlength{\tocsep} 
\JournalInfo{Nov 2014} 
\Archive{} 

\PaperTitle{Temporal patterns behind the strength of persistent ties}%

\Authors{Henry Navarro\textsuperscript{1}, Giovanna Miritello\textsuperscript{1,2}, Arturo Canales\textsuperscript{3}, Esteban Moro \textsuperscript{1}*}
\affiliation{\textsuperscript{1}\textit{Departamento de Matem\'aticas \& GISC, Universidad Carlos III de Madrid, Legan\'es 28911, Spain}}
\affiliation{\textsuperscript{2}\textit{Telef\'onica Digital, Ronda de la Comunicaci\'on s/n, Madrid 28050, Spain}}
\affiliation{\textsuperscript{3}\textit{Telef\'onica I+D, Valladolid 47151, Spain}}
\affiliation{*\textbf{Corresponding author}: emoro@math.uc3m.es} 

\newcommand{\keywordname}{Keywords} 

\Abstract{Social networks are made out of strong and weak ties having very different structural and dynamical properties. But, what features of human interaction build a strong tie? Here we approach this question from an practical way by finding what are the properties of social interactions that make ties more persistent and thus stronger to maintain social interactions in the future. Using a large longitudinal mobile phone database we build a predictive model of tie persistence based on intensity, intimacy, structural and temporal patterns of social interaction. While our results confirm that structural (embeddedness) and intensity (number of calls) are correlated with tie persistence, we find that temporal features of communication events are better and more efficient predictors for tie persistence. Specifically, although communication within ties is always bursty we find that ties that are more bursty than the average are more likely to decay, signaling that tie strength is not only reflected in the intensity or topology of the network, but also on how individuals distribute time or attention across their relationships. We also found that stable relationships have and require a constant rhythm and if communication is halted for more than 8 times the previous communication frequency, most likely the tie will decay. Our results not only are important to understand the strength of social relationships but also to unveil the entanglement between the different temporal scales in networks, from microscopic tie burstiness and rhythm to macroscopic network evolution.}

\begin{document}

\flushbottom 

\maketitle 


\thispagestyle{empty} 

\setstretch{0.92}

\lettrine[nindent=0pt, lines=2]{S}{ocial} networks are dynamic objects, they grow and change over time through the addition of new ties or the removal of old ones, leading to an ongoing appearance and disappearance of interactions in the underlying social structure \cite{Saramaki:2015dy,Holme:2012vb}. Identifying the different mechanisms by which a tie form or decay is a fundamental and challenging question of individual human behavior, but also it can unravel the processes behind group, community and network dynamics that shape our social fabric and, in turn, how that network evolution impact important processes in our society like cooperation \cite{Rand:2011dv}, disease spreading \cite{Holme:2016fm} or information diffusion \cite{miritello2013temporal,PhysRevE.83.045102,karsai2011small}. On the other hand, understanding under what condition a tie is more or less likely to decay may shed light on the circumstances under which an observed interaction can be actually considered a genuine social relationship \cite{Hidalgo:2008p1411, Kossinets:2006p368} and its present and future potential strength in the different processes happening in social networks.

Most of the understanding on the dynamics of tie formation and decay comes from the determination of microscopic factors governing tie formation and persistence \cite{Rivera:2010ci}. In particular a special attention has been given to endogenous factors, i.e. those properties that can be extrapolated from the network itself to predict future tie behavior. Intensity of previous interactions, reciprocity, network proximity, triadic closure or the existence of common friends are not only predictors of tie formation \cite{LibenNowell:2007p2128}, but also of its persistence in the future \cite{Raeder2011245,Hidalgo:2008p1411}. In the context of Granovetter's theory of {\em strength of weak ties}, strong ties are those which are more likely to persist, since they are structurally embedded (common friends) are more intense (number of interactions), while bridges between communities are weak and, as Burt found \cite{DBLP:journals/socnet/Burt00}, they are more likely to decay in the future. Intensity and embeddedness are thus commonly acknowledged as properties behind a strong and/or persistent tie.

Despite these findings, we still have not a comprehensive understanding of what are the main properties of human interaction that make social ties to persist. This is largely due to the lack of quality data: although some online social networks have explicit mechanisms to ``\textit{unfollow}'' (Twitter)  \cite{ICWSM124598} or `\textit{unfriending}'' (Facebook) \cite{Quercia:2012:LFF:2380718.2380751} other users, the access to structural or intensity data in those platforms is limited in those platforms. On the other hand, most studies infer tie decay from absence of tie activity in large databases \cite{Hidalgo:2008p1411, Raeder2011245}. This is a potential problem since, given the large burstiness of human interaction \cite{Barabasi:2005p327,PhysRevE.83.045102}, large inactivity periods could be mistaken as tie decay events. Thus, although previous studies of tie decay agree on the general importance of structural embeddedness, intensity or reciprocity of a tie to predict its future persistence \cite{Raeder2011245,Hidalgo:2008p1411}, they still provide an incomplete picture of what are the main tie properties that make them strong (persistent) and if, as was done in the problem of tie prediction, we can build efficient models based on endogenous properties of ties to predict if a social relationship is bound to decay. 

In this paper we overcome some of these difficulties by studying tie persistence in human communication using a large longitudinal database of 19 month of mobile phone calls. The large duration of the database allow us to accurate assess the presence of a tie by using the method introduce by Miritello {\em et al.} \cite{miritello13} which splits the observation period in different time windows and use each of them to characterize and assess the presence of the tie. But more importantly, having a detailed and large longitudinal database for human communication allow us to characterize better the patterns of communication within a tie and see if temporal properties of human interaction are predictors of tie persistence in the future. Although simple temporal properties have been considered before in the problem of tie prediction \cite{Tabourier:2016ks} and strength estimation \cite{Gilbert:2009:PTS:1518701.1518736,Raeder2011245}, here we show that the persistence of a tie is also encoded in the bursty patterns of communication between people. Furthermore, by building a high accurate predictive model based on different tie features (structural, intensity, intimacy and temporal) we are able to show that temporal properties are indeed as important as intensity and much more than structural properties in predicting tie persistence. Our results show that it is possible to build simple predictive models of network evolution based only on the temporal and intensity properties of the human interaction.



\section{Measuring the strength of a tie}

\begin{figure*}[h!]
\centering
 \includegraphics[scale=0.8]{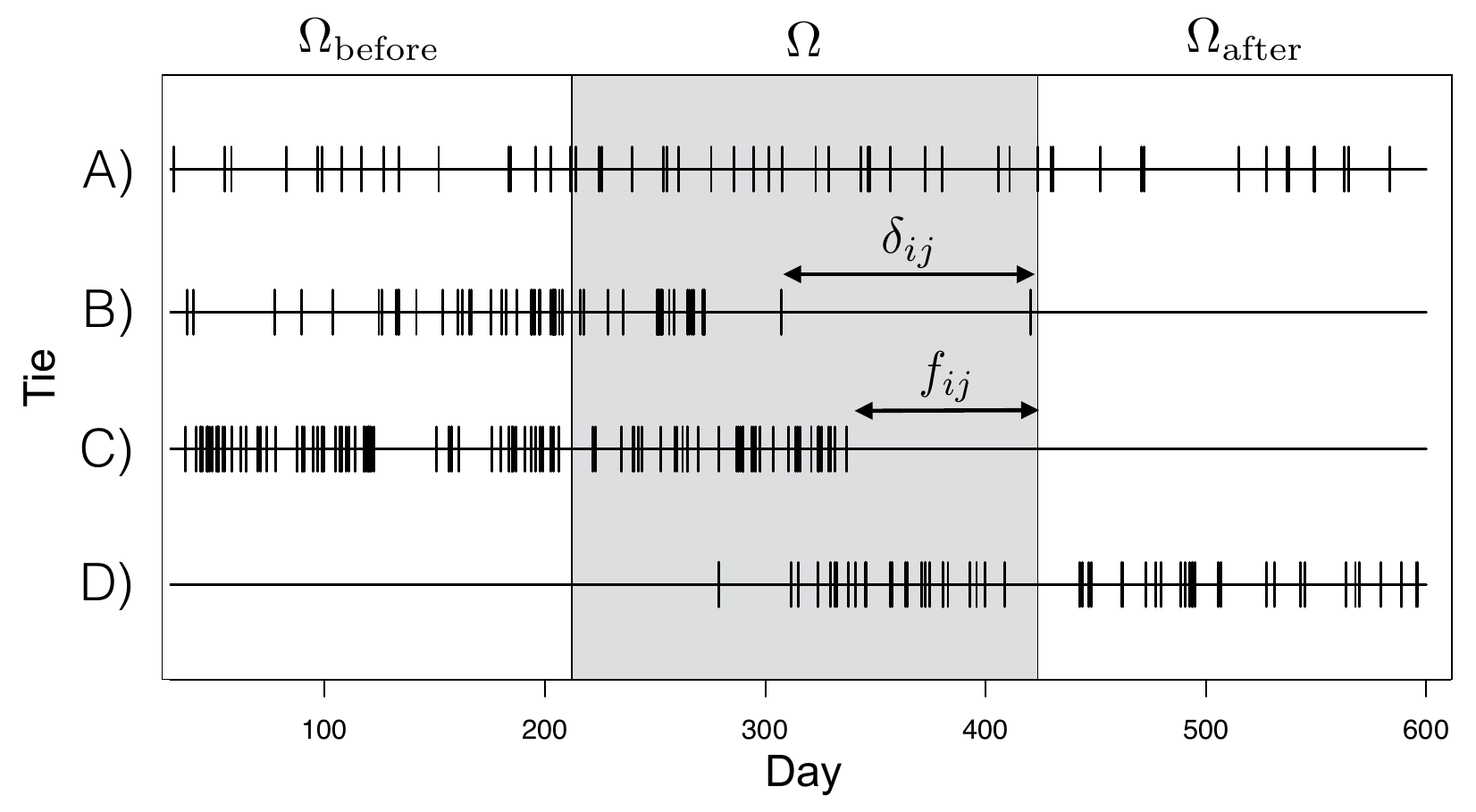}
\caption{{\bf Detecting tie decay and strength.} Definition of observation periods and examples of call activity for 4 given ties. Any vertical segment is a call between the users in a particular tie. Our 19 months database is divided in three periods, where the 7 months in the middle $\Omega$ is our observation period where all the tie features will be measured. The period $\Omega_\mathrm{after}$ is used to asses if ties are persistent, i.e. if there is activity in the tie. For example, ties A) and D) are persistent, while ties B) and C) are said to have decayed in $\Omega_\mathrm{after}$. All ties have similar values of number of calls in the observation period with $w_{ij} \in [30,40]$. We also show specific examples of one inter-event time $\delta_{ij}$ (tie B) and freshness $f_{ij}$ (tie C).}
\label{fig:scheme}
\end{figure*}
To understand that behavior we study a sample of 20000 ties drawn randomly from the \textit{Call Detail Records} (CDR) of 20 million people from a single mobile phone operator over a period of 19 months. As in \cite{miritello13} we divide the time interval in three periods: the 7 months in the middle $\Omega$ define our observation and measurement period for the ties. We only select 13708 ties in which there are at least 5 calls in $\Omega$ between users, and among those calls there has been at least one call in each direction. As in \cite{miritello13}, the first and last periods of 6 months $\Omega_{\mathrm{before}}$ and $\Omega_\mathrm{after}$ are used to assess whether the tie has decayed: since there is no explicit information about whether social interactions stop, we will say that the tie between user $i$ and $j$ has decayed if there are no calls between them in $\Omega_\mathrm{after}$. This functional definition of the existence of a tie underestimates the possibility of having another call after those 6 months, but as it was shown in \cite{miritello13}, only 3\% of ties contain such long inter-event times $\delta_{ij}$ between calls (see figure \ref{fig:scheme}), which shows that our method is subject only to a small error. It is important to understand that since activity within ties is bursty, large inter-events between interactions are likely and thus they might be mistaken as tie decay. In particular, in our database we find that the average time between calls in a tie is $\overline{\delta}_{ij} = 14$ days (with a standard deviation of 18 days), and thus we might get spurious effects if $\Omega_\mathrm{after}$ is of the order of a month, as interactions may fall outside the $\Omega_\mathrm{after}$ period. See the {\em Methods} section for further description of the mobile phone dataset. We have also considered another (smaller) database of Facebook communication through wall posts. Since the results on both databases are similar we discuss here only the mobile phone database and refer to the {\em Methods} section for further details about the Facebook database analysis.

To characterize the strength of the tie we will find those features that can anticipate its persistence. Thus, we will implicitly identify strong relationships with persistency, while weak ties are those more likely to decay. This definition of strength is then a much more functional form of describing its utility in present and future social processes and operationalizes Granovetter's idea that strong ties are those which are more likely to persist.  To describe tie features we will also follow Granovetter's notion of ``strength'' of an interpersonal tie \cite{Granovetter1973}: ``the strength of a tie is a combination of the amount of time, the emotional intensity, the intimacy (mutual confiding), and the reciprocal services which characterize the tie''. Within that framework, we define four categories of tie features: intensity, temporal, structural and intimacy features, and we will try to characterize which ties are the strongest (more persistent) according to these variables. Intensity, frequency and intimacy features will refer to properties of the communication patterns between users, while structural variables are those derived by understanding how the tie is embedded in the rest of the social network. Given the nature of our data, our features will be constructed solely taking into account the information about call events between users.  Our working assumption is that there is enough information in those events to predict the persistence of the tie.

Some of the variables are adapted from previous works both in tie formation and decay prediction \cite{DBLP:journals/corr/WangXWZ14, Quercia:2012:LFF:2380718.2380751, miritello13, Raeder2011245}, but others are introduced for the first time in this work. Specifically we introduce a number of variables that take into account the temporal patterns of the communication between users \cite{Saramaki:2015dy,miritello13}. Contrary to the static and aggregated version of relationships and networks, ties and networks are always evolving: not only communication between users is highly bursty and correlated in time \cite{karsai2011small,PhysRevE.83.045102}, but also the dynamical strategies by which users create and destroy ties are very different \cite{miritello13,Saramaki:2012uj}. The hypothesis we investigate in this paper is whether those patterns convey information about the fate of a social relationships. For example, if the periodicity or burstiness of how two people communicate or if they are involved in very fast social creation and destruction of ties can inform us about the persistence of social ties.

\subsection{Intensity features}
The first group of variables describe the amount of communication between users. Stronger relations imply a more frequent relationship which we can quantify by the number of calls between users $w_{ij}$. This variable is highly heterogeneous in our database in a similar way as other similar works in the literature \cite{Onnela:2007p316} (see Figure \ref{fig:dist}). Specifically we find that the average number of calls is $\overline{w_{ij}} = 76$ while it varies from a minimum of 5 and a maximum of 2468 calls per tie. To take into account this heterogeneity, the rest of the variables we will consider are calculated with respect to that level of activity per tie. For example, we will take the average duration of calls per tie $d_{ij}$ instead of the total duration, because the latter is highly correlated with the number of calls. On the other hand, several works have found that if the tie is highly reciprocal, the relationship is stronger and thus is less likely to decay \cite{Hallinan:1978jb,Hidalgo:2008p1411,Raeder2011245}. Our database contains information about which user initiates the call so we can measure $w^{\rightarrow}_{ij}$, the number of calls from $i$ to $j$ initiated by $i$. Using this, we define the level of reciprocity in between users $i$ and $j$ as 
$$
r_{ij} = \left|\frac{w^{\rightarrow}_{ij}}{w_{ij}}-\frac12\right|.
$$
Note that this variable take values between $0$ and $1/2$. When user $i$ initiates most of the calls in the tie, then $w^{\rightarrow}_{ij} \simeq w_{ij}$ and $r_{ij} \simeq 1/2$. On the contrary, when the number of calls from $i$ to $j$ is equal to the number of calls from $i$ to $j$, we have that $w^{\rightarrow}_{ij} \simeq w_{ij}/2$ and then $r_{ij}=0$. Thus larger values of $r_{ij}$ indicate less reciprocity. 

\subsection{Structural features}
Formation and decay of a tie is also related with the social structure around it. People tend to form groups and in particular, people tend to form relationships with friends of friends (triadic closure) which leads to high clustering around a tie \cite{Rivera:2010ci}. This is the reasoning behind Granovetter's influential ``strength of weak ties'' argument which implies that not also structural embedded ties are more likely to arise in a social network but they are also more persistent, a result corroborated by Burt in different works \cite{DBLP:journals/socnet/Burt00,Burt2002333}. Although there are many metrics to quantify embeddedness of a tie within the social network, we will use the topological overlap $o_{ij}$ defined as the fraction of neighbors of $i$ and $j$ which are shared \cite{Onnela:2007p316}. Specifically, 
\begin{equation}
o_{ij}=\frac{\big|n_i \cap n_j\big|}{\big|n_i \cup n_j\big|},
\end{equation}
where $n_i$ and $n_j$ are respectively the set of neighbors of the two nodes $i$ and $j$ and $|n_i|$ indicates the number of them. Note that, this variable takes value between 0 and 1, because if $i$ and $j$ have no common neighbors, then $o_{ij}$ will take value $0$. On the contrary, if $i$ and $j$ call to the same circle of id's $o_{ij}$ will take value $1$. The topological overlap is then a variable measuring the (normalized) number of ``common friends'' between two nodes.

The topological overlap is a particular way to measure the structural similarity of users. Another metric we will consider is the disparity (or similarity) between the connectivity of users. In particular, if $k_i$ and $k_j$ are the number of neighbors of $i$ and $j$ we will construct the geometric mean of connectivity $k_{ij} = \sqrt{k_i\ k_j}$. This geometrical mean takes small values if connectivity of users is very different and large values if connectivity is similar (for same levels of connectivity). This variable is also introduced to take into account the effect of the different importance of a tie for the users involved in the relationship. If $k_{ij}$ is small, the tie between $i$ and $j$ is important for both or one of them, while if $k_{ij}$ is large, then it is just another tie among the many they have. Variations of structural connectivity similarity have been considered in other works studying tie strength and dynamics \cite{Raeder2011245,Gilbert:2009:PTS:1518701.1518736}.

\subsection{Intimacy features}
Following Granovetter's hypothesis of a strong tie, the intimacy (mutual confidence) between two nodes could provide a better characterization of the tie and allow a more accurate prediction of its dynamics. As opposed to other studies in social networks \cite{Gilbert:2009:PTS:1518701.1518736} our mobile phone database does not contain any information about the context and content of the call. Thus we quantify the mutual confidence by the day or hour when the call is made and specifically, we consider the fraction of calls within a tie that are made after 8pm and during the weekend, $\mu^{int}_{ij}$. As was shown recently, calls made in the evening and at night are typically focused on a small number of emotionally intense relationship \cite{aledavood2015daily} and thus, quantifying the amount of communication happening at that time of the day can give us a proxy for intimacy. 

On the other hand, difference of demographic characteristics of users have an impact in tie dynamics. For example, the temporal communication patterns formed by groups of male or female are different \cite{onnela2014using}, and those patterns can be associated with the different preference strategies of both sexes across the lifespan \cite{palchykov2012sex}. To quantify those relationship preferences, we consider the age and gender difference between the users participating in a tie. Age difference $age_{ij}$ is measured as the absolute value of the difference in years while gender difference is a dichotomous variable where $gender_{ij} = 1$ if both users have same gender and $gender_{ij} = 0$ if they are different.

\subsection{Temporal features}
Finally we characterize the temporal patterns within and around the tie. Since communication within the tie is very heterogeneuous (see figure \ref{fig:scheme}), we want to understand whether that heterogeneity might reveal something about the persistence of the tie. The first variable we consider is the {\em freshness} of the tie $f_{ij}$, i.e. the time since the last call between $i$ and $j$ at the end of $\Omega$ \cite{Gilbert:2009:PTS:1518701.1518736,Raeder2011245}. As before, since activity within ties is very heterogeneous, we consider the relative {\em freshness} as the relative time elapsed from the last call compared to the typical time between calls in the tie $\hat f_{ij} = f_{ij} / \overline{\delta}_{ij}$ where $\overline{\delta}_{ij}$ is the average inter-event time between calls. At the same time we also consider the age of the tie as the time of the first call between users in our database $t^{min}_{ij}$ measured in days.

Another tie we consider is the burstiness of the communication patterns. More regular communication patterns could be related to stronger and more intimate relationships and thus, less bursty communication patterns might persist more. Although there are many ways to characterize burstiness of events \cite{goh2008burstiness}, we will use two simple metrics. The first one is the coefficient of variation of the inter-event times $cv_{ij} = \sigma_{ij}/\overline{\delta}_{ij}$, where $\overline{\delta}_{ij}$ is the average inter-event time between two calls and $\sigma_{ij}$ is their standard deviation. If $cv_{ij} \gg 1$ then communication is very bursty, with large untypical periods of time in which users didn't communicate (see for example tie B in figure \ref{fig:scheme}), while if $cv_{ij} \ll 1$, communication was very regular, happening almost at the same time intervals (see tie A in figure \ref{fig:scheme}). The value $cv_{ij} = 1$ correspond to the Poissonian homogenoeus case in which inter-event times are distributed randomly along the $\Omega$ period \cite{goh2008burstiness}. Another way to characterize the burstiness is to quantify how many communication events happened in bursts or rapid consecutive successions of calls (we will call them {\em chats}) \cite{karsai2012universal,PhysRevE.83.045102}. To do that we calculate the fraction of calls $\mu^{chats}_{ij}$ that happened only with 5 minutes difference between them.

\begin{table*}[h!]
\caption{Features of ties between user $i$ and $j$ considered to characterize the strength (persistence) of the ties, including their (normalized) complexity measured in computational time.}
\begin{center}
\begin{tabular}{lclc}
\hline \hline
Type & Feature & Description & Computational complexity\\
\hline
Intensity & $w_{ij}$ & Total number of calls  &	1.00\\
Intensity & $d_{ij}$ & Average duration of calls & 1.00 \\
Intensity & $r_{ij}$ & Reciprocity of calls & 1.12 \\
\hline
Structural & $o_{ij}$ & Topological overlap & 1.82 \\
Structural & $k_{ij}$ & Connectivity diversity & 1.33 \\
\hline
Intimacy & $\mu^{int}_{ij}$ & Fraction of calls after 8am and weekends & 1.05\\
Intimacy & $age_{ij}$ & Age difference in years & 0.18\\
Intimacy & $gender_{ij}$ & Gender difference & 0.15 \\
\hline
Temporal & $\hat f_{ij}$ & Relative freshness & 1.01\\
Temporal & $t^{min}_{ij}$ & Age of tie (in days). & 1.01\\
Temporal & $cv_{ij}$ & Inter-event time coefficient of variation & 1.11\\
Temporal & $\mu^{chats}_{ij}$ & Fraction of consecutive calls (5 mins.) & 1.31\\
Temporal & $a_{ij}$ & Users' Activity diversity & 1.21\\
\hline
\hline
\end{tabular}
\end{center}
\label{tabla1}
\end{table*}

Finally, another reason why a tie decays is simply because users involved in the tie have very different dynamical social strategies. As was found in \cite{miritello13} humans constantly create and destroy ties and they have different strategies to do that. While some individuals do create and destroy a lot of ties ({\em explorers}), others tend to maintain in time their social circle ({\em keepers}). If both users in a tie are explorers, the probability for the tie to decay is high. To measure how dynamical are the strategies of users in a tie we consider $a_i$, the number of ties created by user $i$ in period $\Omega$. As in \cite{miritello13} we say that a tie is created in $\Omega$ if there is no call between users in $\Omega_\mathrm{before}$. The ratio between the number of created ties and the total number of ties $a_i/k_i \in [0,1]$ describe how frequent user $i$ changes her social neighborhood. If $a_i/k_i \simeq 1$ it means that most of the ties of user $i$ where created during $\Omega$ (i.e. the user {\em social explorer}), while if $a_i/k_i \ll 1$ most of the ties are stable ({\em social keeper}). To characterize how dynamical are the strategies of both $i$ and $j$ we consider the geometrical mean of
\begin{equation}
a_{ij}=\sqrt{\frac{a_i}{k_i}\cdot\frac{a_j}{k_j}}.
\end{equation}
If both $i$ and $j$ are explorers, $a_{ij} \simeq 1$ and the tie is more likely to decay since it connects users with highly dynamical social strategies, while if they are both keepers, $a_{ij} \simeq 0$ and the tie most likely will persist.\\

Table \ref{tabla1} summarizes the features considered to assess the strength of persistent ties. Before constructing our models and because of the large heterogeneity found in connectivity, activity and burstiness across ties in social networks, we scale and normalize our variables before using them in a model. For example, we consider $\log w_{ij}$ instead of $w_{ij}$ since the distribution of number of calls per tie is heavy skewed in mobile phone databases \cite{Onnela:2007p316}. On the other hand the burstiness within ties make variables like $cv_{ij}$ or $\hat f_{ij}$ also very heavy-tailed across our dataset. Thus we also use a logarithmic scaling for them. Although they are logarithmically scaled, in the rest of the paper we denote them by its original name for sake of clarity, unless were numerical values are given (for example in figure \ref{fig:corr1}). Finally, since the correlation between the variables is small, we keep all features in our analysis (see {\em Methods} section to learn about the preprocessing and selection of variables).

\section{Results}

\begin{figure*}[th!]
\centering
 \includegraphics[scale=0.75]{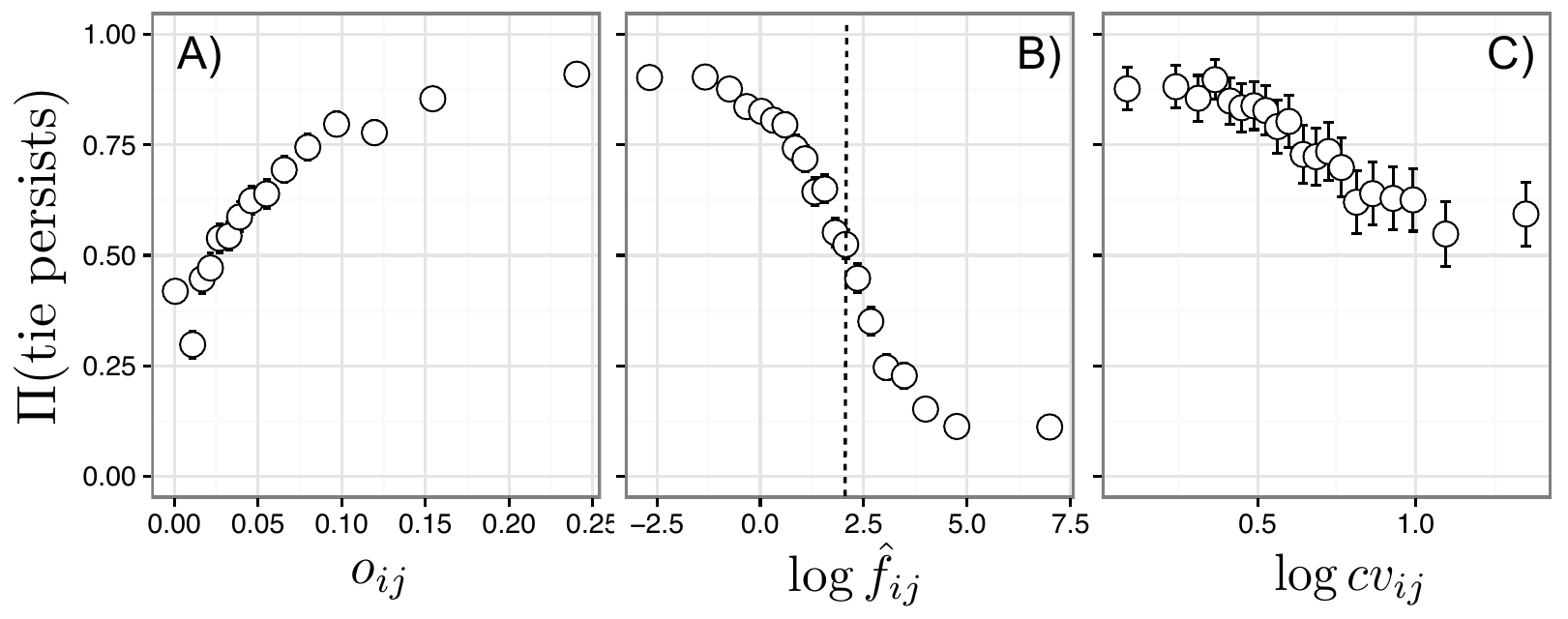}
\caption{{\bf Impact of some features on tie persistence.} Conditional probability for the tie to persist as a function of the different variables. In C) only ties with $w_{ij}\in [20,50]$ are considered. The vertical line in B) indicates the critical relative freshness $\hat f_{ij} = 8.33$, where $\Pi = 1/2$. }
\label{fig:examples}
\end{figure*}

A simple inspection of how persistence depends on some tie features corroborates some results found in the literature. For example, as Burt found in \cite{DBLP:journals/socnet/Burt00} we observe that weak ties with small topological overlap have more probability to decay (see figure \ref{fig:examples}A), i.e. bridges are more likely to decay while persistent ties are those embedded within communities. Note that this effect can have a 50\% change in probability from ties with no overlap $o_{ij} = 0$ to the largest overlap observed in the database $o_{ij} \simeq 0.5$. Similarly to \cite{Gilbert:2009:PTS:1518701.1518736} we find that the time since the last communication also reveals how likely is to observe activity in the tie again: most recent activity implies that the tie will persist in the future (see figure \ref{fig:examples}B). Finally, we find that some temporal features are strongly correlated with tie persistence. For example in figure \ref{fig:examples}C we find the interesting result that more bursty communication within a social tie is correlated with tie decay.

Although this individual results demonstrate the potential predictive power of our tie features to get a complete picture of tie persistence we build a predictive model of tie decay based on {\em all} the features introduced in the last section. We define two different prediction models depending on the reference frame used to characterize tie strength. In the first one ({\em Model 1}) we used a fixed reference frame for all ties, namely we try to predict if the tie decays in $\Omega_\mathrm{after}$ by observing its features along $\Omega$. Although this is the traditional setting for tie persistence prediction, the features calculated during $\Omega$ might be impacted by the fact that the tie decayed early in the interval $\Omega$ (see for example tie C in figure \ref{fig:scheme}). If this happens, variables like the number of calls, their duration, or the structural overlap are going to be naturally smaller just because the tie decayed earlier. By including all those early decay events, {\em Model 1} is going to incorporate some information about what happens after the tie decays, making it difficult to disentangle what part of the prediction power comes from properties of the tie before or after it decays. For this reason we will build another predicting model {\em Model 2} in which we will only consider those ties that have a call within the last two weeks of $\Omega$. This way we will use a relative reference frame in which we want to understand what properties of an existing tie have more impact in its immediate future stability. Both models are important to understand the dynamics of a tie, its stability, and in general, the evolution of networks. But {\em Model 2} might give a more direct understanding of what defines a strong social relationship without requiring a long time interval to observe if there was a significant decay in the activity of the tie.

To predict tie persistence we build a classification model using simple logistic regression (LogR) models where the positive class is that the tie persists, that is, that we observe at least a communication event in $\Omega_\mathrm{after}$. We use a train dataset using 75\% of our ties and 10-fold cross validation to fit the probability for a tie to persist using the inverse logit function
$$
\Pi(\mathrm{tie}\ ij\ \mathrm{persists}) = \frac{1}{1+e^{-\beta_0 - \sum_{l=1}^n \beta_l x_l}}
$$
where $x_l$ are the features introduced in the last section and $\beta_l$ are the coefficients obtained in the fit. Note that positive values of $\beta_l$ indicate that the variable $x_l$ as a positive effect in the persistence of the tie: larger values of $x_l$ increase the probability for the tie to persist. The performance of the model is measured using the rest 25\% of our ties, achieving values around 0.8 for its accuracy, sensitivity and specificity, showing the good balance of our model detecting both classes (persistent and decaying ties). Details of how the predicting model was constructed can be found in the {\em Methods} section.

The results for the different models are presented in table \ref{table:glmmodels}, where we can see that, as expected, variables like the number of calls $w_{ij}$, mean duration $d_{ij}$ or topological overlap $o_{ij}$ have a positive effect in tie persistence \cite{Hidalgo:2008p1411,Raeder2011245}: the larger they are the more likely the tie will persist in the future. Interestingly, the same happens with gender difference: ties that tie individuals with equal gender are more persistent than those between persons of different gender, a reflection of the same-gender homophily previously found in the most stable relationships \cite{palchykov2012sex}. However, other well studied variables like reciprocity, connectivity diversity o age difference seem not to be important for tie persistence. 

Temporal variables play a major role in the models. Specifically, newer ties (smaller $t^{min}_{ij}$) are more likely to be observed in the future which might reflect the fact that newly stablished ties take some time to decay. But more importantly, in {\em Model 1} the persistence of the tie is highly determined by the (relative) freshness $\hat f_{ij}$, i.e. how much time has passed since the last communication between users: as we can see, the coefficient is negative, which means that larger times since the last communication mean smaller probability for the tie to persist. Other temporal variables like the coefficient of variation and number of chats have some impact on the persistence of the tie. For example, larger number of rapid consecutive calls (larger $\mu_{ij}^{chats}$ or more regular patterns (smaller $cv_{ij}$) yield to better stability of ties, an interesting result that shows that high frequency patterns of communication between users also encode some information about how strong is the tie. Finally, the coefficient for $a_{ij}$ is negative, i.e, if users participating in the tie have more {\em explorer} behavior, the tie has lower probability to persist.

\begin{figure*}[h!]
\centering
 \includegraphics[scale=1.1]{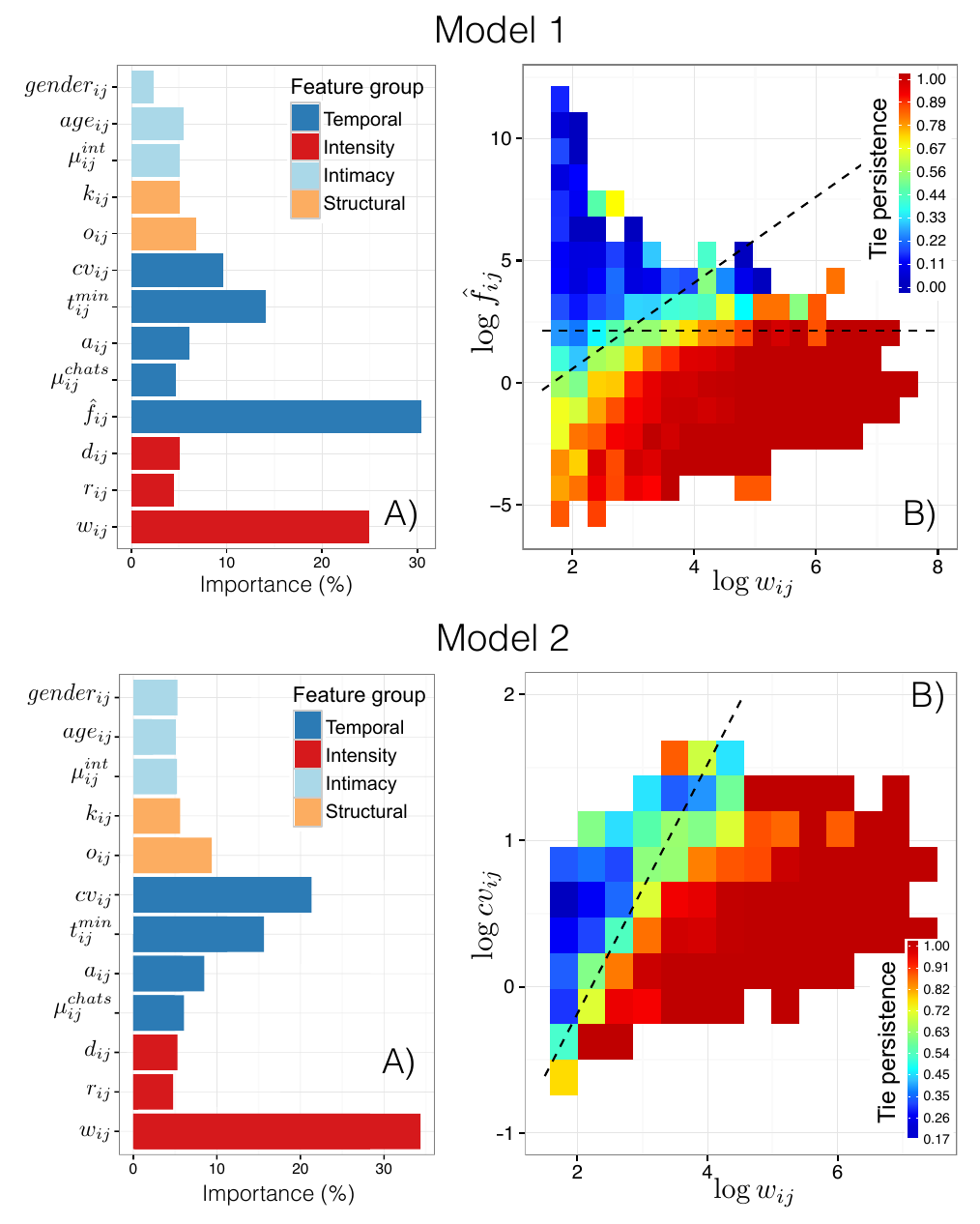}
\caption{{\bf Feature importance in tie persistence.} A) Importance of the variables in the full model in table \ref{table:glmmodels}. Importance is measured as the normalized \% of the t-statistics for each model parameter. B) Density plot of the average persistence of ties as a function of the two most important variables in A), namely, normalized freshness and total number of calls. The dashed line shows the $1/2$ probability for the simplified models in table \ref{table:glmmodels}. }
\label{fig:corr1}
\end{figure*}

However, not all the variables have equal importance in the persistence model. All together, temporal variables are the most important variables in the model: they amount around $\sim60$\% of the importance in our predictive model (see figure \ref{fig:corr1}), while intensity variables giving $\sim30\%$ of the importance and finally structural and intimacy variables representing less than $\sim 10\%$ (each) of the model importance. The relative small importance of well studied properties like the topological overlap $o_{ij}$ could be due to the Granovetter effect, i.e. the fact that since $o_{ij}$ and $w_{ij}$ are moderately correlated yields $o_{ij}$ to have less importance in the model, since its effect is already included in $w_{ij}$. As we can see in figure \ref{fig:corr1} it is remarkable to see that just two variables (number of calls $w_{ij}$ and relative freshness $\hat f_{ij}$ or coefficient of variation $cv_{ij}$) have most of the importance in model to the point that a simplified model based on only those two variables achieve similar levels of performance (see Table \ref{table:glmmodels}). In the case of {\em Model 1}, actually, just the number of calls or the relative freshness achieve a high accuracy (78\%), a result that can be shown graphically in figure \ref{fig:corr1} where the diagonal dashed line corresponds to the $\Pi = 1/2$ probability. Interestingly, similar level of accuracy is found for the really simple model based on just the relative freshness (horizontal line in figure \ref{fig:corr1}). In that case $\Pi = 1/2$ corresponds to a critical relative freshness of $\hat f_{ij} = 8.33$ so ties with larger/smaller values have less/more than 50\% probability to persist. This result shows that ties in which the natural rhythm of their communication is halted have more probability to decay. Specifically we found this happens when the last interaction between users happened at least 8.33 times their typical inter-event time. As an example, if two users typically called themselves each day in the past and more than 2 weeks have elapsed since their last communication, the tie might have decayed. 

In the case of {\em Model 2} we also find that intensity and temporal properties are the most important variables to explain tie persistence giving respectively $\sim 43\%$ and $\sim 50\%$ of the importance of the model, as we can see in figure \ref{fig:corr1}. But also we can explain most of its accuracy by a simplified model in which only the number of calls and the coefficient of variation are considered, see diagonal dashed line in figure \ref{fig:corr1}. The strong importance of $cv_{ij}$ in the model signals a very interesting fact: for a fix given level of activity $w_{ij}$, ties which are more bursty (high $cv_{ij}$) have more probability to decay. This finding suggest that special attention paid by users to maintain a periodic communication might be an indication of a more strong and persistent relationship, while highly bursty and heterogeneous call patterns might be a sign of an informal or casual relationships that could decay in the near future. 

Another dimension controlling the effectiveness of the different variables in a predictive model is their complexity. While some of the variables are easy to compute for a given dataset, other features like topological overlap $o_{ij}$ or users activity diversity $a_{ij}$ are very complex, i.e. they need larger computational time. Table \ref{tabla1} shows the computational time (in seconds) to compute each tie feature normalized to the time it takes to compute $w_{ij}$. As we can see structural features like topological overlap or connectivity diversity are very costly (up to 1.82 times the total number of calls), while temporal features are cheaper to compute. This result, together with the low predictive power of traditionally considered variables like $o_{ij}$ or $r_{ij}$ shows that temporal features could be much more efficient to detect and predict future tie persistence in a social network.

\begin{table*}[!ht]
\centering
\caption{Regression results for the tie persistence using generalized linear models for the two prediction models. Coefficients are shown with uncertainties (standard errors) in parentheses. Model {\em Full} include all the features described in the text, while model {\em Simplified} only includes the most important two features.}
\begin{tabular}{ c | c  c c | c c | c}
				& 		\multicolumn{3}{c}{\textbf{Model 1}} & \multicolumn{2}{c}{\textbf{Model 2}} & \textbf{Time}\\ 
\textbf{Feature} & Full & Simplified & Simplified' & Full & Simplified  \\ \hline
\hline
$w_{ij}$ 		& $1.328^{***}$	& $1.146^{***}$ &	& $2.428^{***}$	& $2.041^{***}$	& 0.9956	\\ 
				& (0.041) 		& (0.029)		&	& (0.155)		& (0.105)	 \\
$d_{ij}$ 		& $0.055^{*}$	& 				&	& $0.029^{}$	& 	0.9956 & 			\\
				& (0.025) 		& 				&	& (0.071)		& 				\\
$r_{ij}$ 		& $-0.038$ 		&				&	& $-0.059$		&				\\
				& (0.023) 		& 				&	& (0.066)		& 			\\
\hline
$o_{ij}$ 		& $0.202^{***}$	& 				&	& $0.119^{}$	& 			\\
				& (0.033) 		&				&	& (0.104)		& 				\\
$k_{ij}$ 		& $0.005$		&  				&	& $0.047^{}$	& 				\\
				& (0.025) 		&				&	& (0.069)		& 			\\
\hline
$\mu_{ij}^{int}$ & $0.059^{*}$	&  				&	& $0.098$		& 		\\
				& (0.023) 		& 				&	& (0.065)		& 		 \\
$age_{ij}$ 		& $0.015$		&  				&	& $0.013$		& 			\\
				& (0.023) 		& 				&	& (0.072)		& 		\\
$gender_{ij}$ 	& $0.107^{***}$	&  				&	& $0.097$		& 		\\
				& (0.023) 		& 				&	& (0.069)		& 	 \\
\hline
$\hat f_{ij}$ 	& $-1.275^{***}$& $-0.652^{***}$& $-1.483^{***}$	&				& 	 \\
				& (0.033) 		& (0.015)		& (0.029)	&				& 	\\
$t_{ij}^{min}$ 	& $-0.293^{***}$&  				&	& $-0.434^{***}$& 	\\
				& (0.028) 		& 				&	& (0.085)		& 	\\
$cv_{ij}$ 		& $-0.289^{***}$	&  				&	& $-0.732^{***}$&  $-2.394^{***}$\\
				& (0.026) 		& 				&	& (0.079)		&  (0.224) \\
$\mu_{ij}^{chats}$ & $0.141^{***}$& 			&	& $0.164^{*}$	& 			\\
				& (0.027) 		& 				&	& (0.071)		& 				\\
$a_{ij}$ 		& $-0.122^{***}$&  				&	& $-0.210^{*}$	& 				\\ 
				& (0.028) 		& 				&	& (0.083)		& 				\\ \hline
Constant 		& $0.624^{***}$ & $-1.917^{***}$&$0.347^{***}$	& $1.001^{***}$	& $-4.525^{***}$ \\
				& (0.025) 		& (0.082)		&(0.020)	& (0.092)		& (0.265) \\
\hline 
Number of points & 13708		& 13708			& 13708	& 1722			& 1722 \\
\hline
\textbf{Performance} & 			& 				& 	&				& 				 \\ \hline
Accuracy 		& 0.801 		& 0.785			& 0.747	& 0.798			& 0.799				 \\
Sensitivity 	& 0.824 		& 0.819			& 0.837	 & 0.814			& 0.802				\\ 
Specificity 	& 0.771 		& 0.741	 		& 0.626	& 0.776			& 0.797					\\
\hline
\multicolumn{5}{l}{\textit{Note:} $^{*}p<0.1$;  $^{**}p<0.05$; $^{***}p<0.01$}  \\
\end{tabular}
\label{table:glmmodels}
\end{table*}

\section{Discussion}
Human behavior display very different temporal patterns due to many constrains like circadian rhythms, cognitive limits or finite capacity to perform tasks \cite{Saramaki:2015dy,Aledavood:2015jf}. Since most of those constrains are common to human nature, those patterns show also a large degree of universality across individuals. Interestingly, deviations from universal rhythms can inform us about changes of behavior related to, for example, unemployment \cite{10.1371/journal.pone.0128692}, health conditions \cite{Madan:2012vk}, or crowd events \cite{Botta:2017ct,Dong:2015ht}. Along this line, our research also shows that future network dynamics is encoded in the relative properties of the temporal patterns of communication between individuals and that those temporal properties have more predicting power than structural, intensity or intimacy features of the communication. Specifically, we found if tie activity is not observed for more than 8 times its typical inter-event time, the tie has a great probability to decay, a result that indicates that each tie as a natural rhythm and that when communication is halted for a long time it will probably decay. More importantly, although recent research has found that burstiness affect a large number of human activities and some explanations have been given to explain its universality \cite{Barabasi:2005p327}, our results show that relative burstiness could be also related to the weakness of ties and that those ties that show excessive burstiness might decay in the future. Since burstiness in ties slows down information spreading \cite{PhysRevE.83.045102}, we have found that more bursty ties are not only weaker to transmit information, but also they are more prone to disappear, making them extremely fragile for the structural and functional processes happening in social networks. 

Our analysis reveal that there is a large entanglement between the different time scales present in social networks and that analyses based on pure structural static features of human relationships might give a partial and biased description on the evolution of our communities, groups and societies \cite{Saramaki:2015dy,Ubaldi:2017db}. For example, short time scales (minutes, time between calls in a tie) seem to foresee the decay of ties in the future (month time scale). More importantly, it seems that temporal properties of ties are better and more efficient descriptions of the social strength than structural features, which will allow faster and simpler detection of changing events in the topology of social networks. In fact we find that structural features like topological overlap play a minor role in our model. This is probably the result of the moderate correlation between the strength and embeddedness in social networks (the Granovetter effect \cite{Granovetter1973}), but also shows that a better picture of strong/persistent ties can be obtained just by looking at temporal and intensity features of social relationships. Our results are in line with recent measures of strength of social ties in social media \cite{Gilbert:2009:PTS:1518701.1518736} where structural variables account only for 4.5\% of tie strength. The same small impact of common friends was found in detecting tie persistence \cite{Raeder2011245}. This body of research and our results seem to imply that, although in the absence of tie activity social structure could be a good (and probably the only) predictor of the formation of a tie \cite{LibenNowell:2007p2128}, once the tie is formed its strength or persistence is immediately encoded into the intensity and temporal features of the interaction.

Finally, a possible explanation of our results might be in the way people share their attention and time over their relationships, giving more frequent and more regular attention to stronger ties than to the weak ones. As we know, humans are bounded by time, money or cognitive limits and   they make decisions to share their time across tasks (including the social ones) causing irregular (bursty) activity. Our findings show that strong and persistent ties suffer less from those bursty patterns, indicating that those ties might have different weight in evaluating how to share our time \cite{Weng:2015wd,Saramaki:2012uj}. We hope our results will help future research to identify better what is the origin of the temporal signs of strong and/or weak ties in social networks.

\section{Methods}

\subsection{Mobile phone data}
As in \cite{miritello13} the data used in this study has been obtained from the Call Detail Records database of a unique mobile phone operator in a single country. We focused exclusively on voice calls records, filtering out short text messages, multimedia messages and operator calls. Each subscription is anonymized such that it is not possible to recover personal information of the users. We filtered out all the incoming or outgoing calls that involve other operators due to the partial access we have to the activity of other providers. To avoid business-like subscriptions, which usually appear as users with a huge number of connections and calls never returned, we only retain ties which are reciprocated, which leads to the removal of about the 50\% of the total links in our database. This restriction also eliminates calls to wrong numbers, telemarketing-type calls, customer service lines, etc. Within this approach, we neglect the directionality of links and consider a call from user $i$ to user $j$ equivalent to a call from $j$ to $i$. The resulting mobile graph contains the communication of 20 million users. 

Since we are interested only in tie dynamics between individuals, we have to take into account the problem of subscription and churn of users in our database. For example, subscription of a new user and its communication with other users in our database results into formation of many new ties for the new subscriber. The same would happen for the decay of ties of a subscribe that churns from the company. To mitigate this problem, we only keep active users in our data set: in particular, we only consider those users who are involved (as calling or as called party) at least in one communication event in each of the three subintervals in the 19 months and also if they are present in the database at least one month before $\Omega$ and are still active one month after $\Omega$. This latter filter prevents spurious effects in the analysis of tie dynamics just because individuals subscribe/unsubscribe just before/after $\Omega$; for example, we could have observed an apparent rapid growth of their social network at the beginning of the observation window or a fast dissolution at its end \cite{miritello2013temporal}. This results in the removal of about the 17\% of nodes and the 37\% of reciprocated links within $\Omega$.

To disentangle the dynamics of ties creation/removal from their call activity, we split the 19 months in 3 subintervals. We have only considered the evolution and properties of the ties within $\Omega$, the 7 months observation period in the middle, using the last 6 months to assess the persistence of the tie. Since we are interested only in tie dynamics between individuals, we have to take into account the problem of subscription and churn of users in our database. For example, subscription of a new user and its communication with other users in our database results into formation of many new ties for the new subscriber. The same would happen for the decay of ties of a subscribe that churns from the company. To mitigate this problem, we only keep active users in our data set: in particular, we only consider those users who are involved (as calling or as called party) at least in one communication event in each of the three subintervals in the 19 months and also if they are present in the database at least one month before $\Omega$ and are still active one month after $\Omega$. This latter filter prevents spurious effects in the analysis of tie dynamics just because individuals subscribe/unsubscribe just before/after $\Omega$; for example, we could have observed an apparent rapid growth of their social network at the beginning of the observation window or a fast dissolution at its end \cite{miritello2013temporal}. This results in the removal of about the 17\% of nodes and the 37\% of reciprocated links within $\Omega$. In our analysis we have considered 20000 random ties from the remaining reciprocated links of the mobile phone graph that have some activity in $\Omega$.

\subsection{Predicting models}\label{predictingmodels}
To predict tie decay/persistence we have used a simple logistic regression model where the positive class is that the tie persists, that is, that we observe at least a communication event in $\Omega_\mathrm{after}$. Since the fraction of ties that decay is small (only 20\% in our sample) our classification problem is slightly unbalance, which might cause problems when training our algorithm. To palliate this problem we use the SMOTE algorithm \cite{chawla2002smote} to generate synthetic cases for the minority class (decay) so that the number of ties that persist and decay is around 50\%. We split our new dataset into a train and test samples which contain respectively 75\% and 25\% of the ties and use 10 fold cross-validation to train the model with Area Under the Curve (AUC) as the performance metric. Performance of the model is evaluated using the 25\% test sample of the data.

\begin{figure*}[h!]
\centering
 \includegraphics[scale=0.70]{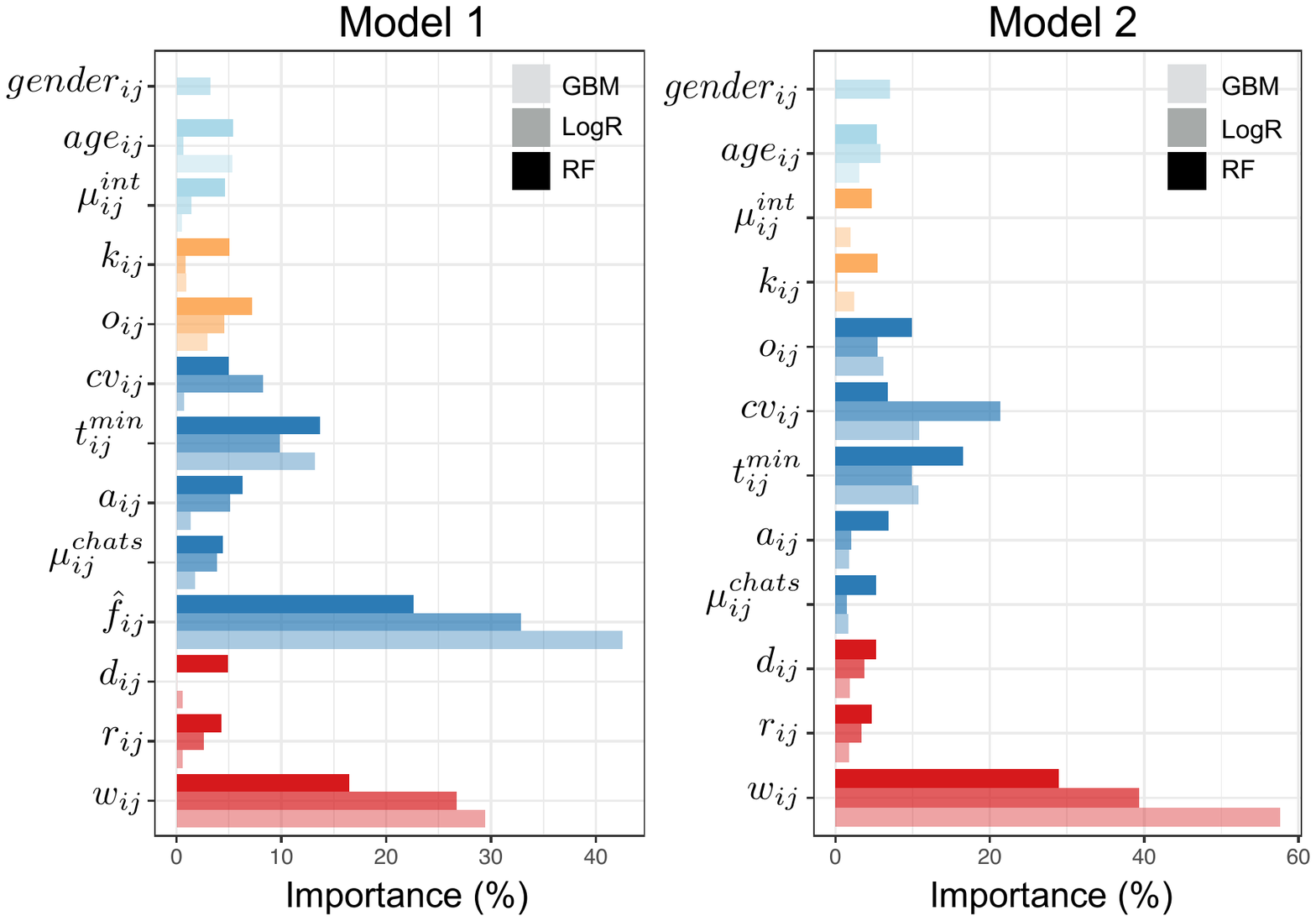}
\caption{{\bf Comparison with other algorithms.} Comparison of the importance of the variables for the two models using different algorithms, namely Random Forests (RF), Logistic Regression (LogR) and Generalized Boosted Regression Models (GBM). As we can see models agree on the relative importance of individual and group variables. Color of the variables correspond to the intimacy (light blue), intensity (orange), temporal (blue) and intensity (red) groups.}
\label{fig:othermodels}
\end{figure*}

To test that our results are not due to the particular algorithm used to predict tie persistence, we have also used other predicting models for this two-classes classification problem. Specifically we have used Random Forests (RF) and Generalized Boosted Regression Models (GBM) \cite{friedman2001elements}. As we can see in figure \ref{fig:othermodels} results are very similar for the different importance of variables. However accuracy is bigger in RF (91\%) and GBM (85\%) when compared with the logistic regression (LogR). This comparison shows that our results do not depend on the actual algorithm used to build the predictive algorithm and that the importance of temporal variables is a genuine finding in our data.

\subsection{Normalization and selection of tie features}
\begin{figure*}[h!]
\centering
 \includegraphics[scale=0.75]{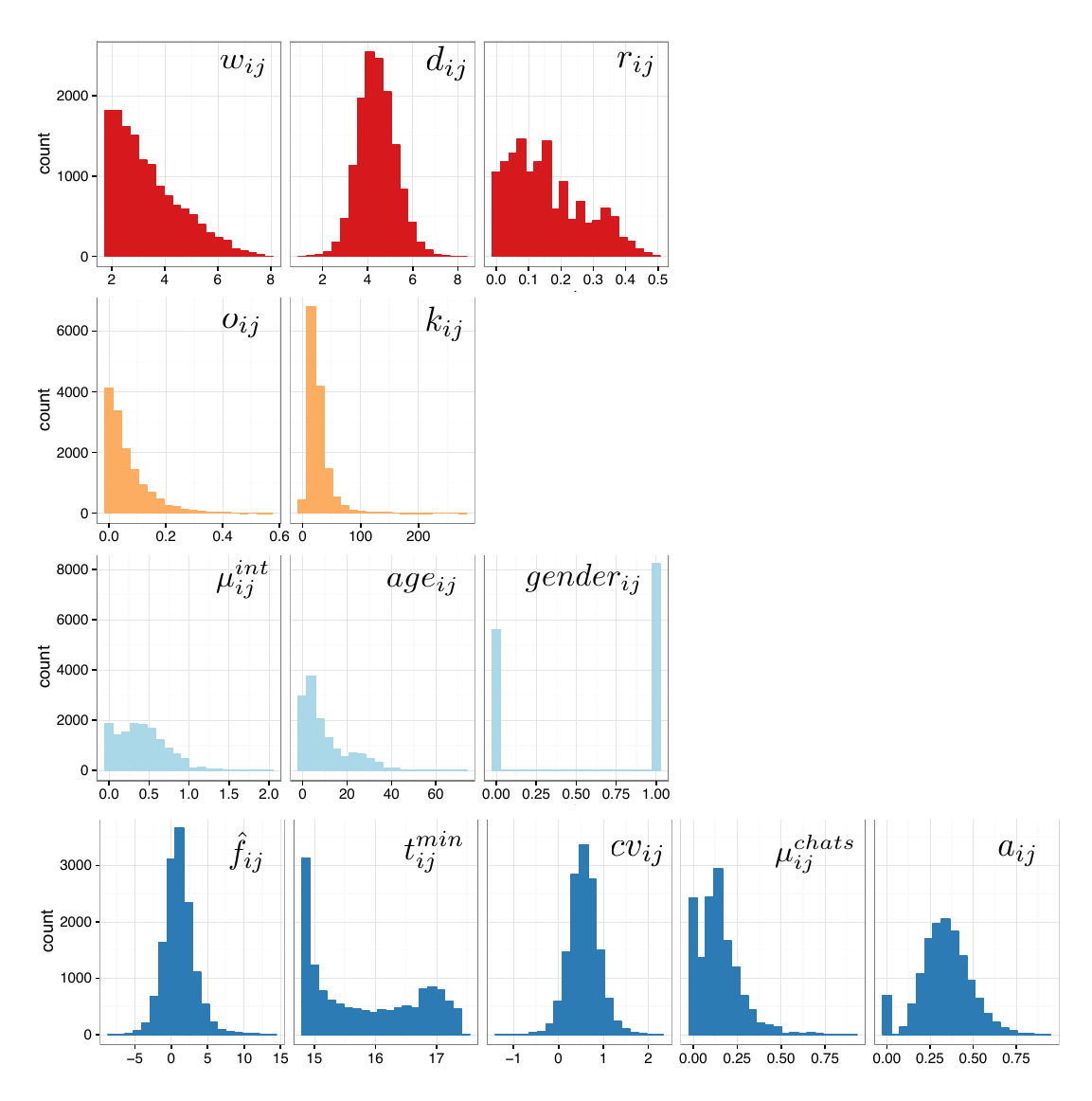}
\caption{{\bf Distribution of features.} Histograms of the different features considered in our models. Each row of histograms correspond to a different group of features: intensity, structure, intimacy and temporal features from top to bottom. }
\label{fig:dist}
\end{figure*}

In the logistic regression classifier is common to implement some kind of normalization of variables through transformations. This is specially important when variables have highly skewed distributions as is typically found in variables describing human activity and behavior. In our case  variables like the intensity $w_{ij}$, average duration $d_{ij}$, relative freshness $\hat f_{ij}$, time since the first call $t_{ij}$ and coefficient of variation $cv_{ij}$ are heavy-tailed distributed and thus we have log-transformed them before using them in our models. As we can see in Figure \ref{fig:dist}, after this transformation, the histogram of the main variables used in our models is more homogeneous.

\begin{figure*}[h!]
\centering
 \includegraphics[scale=0.75]{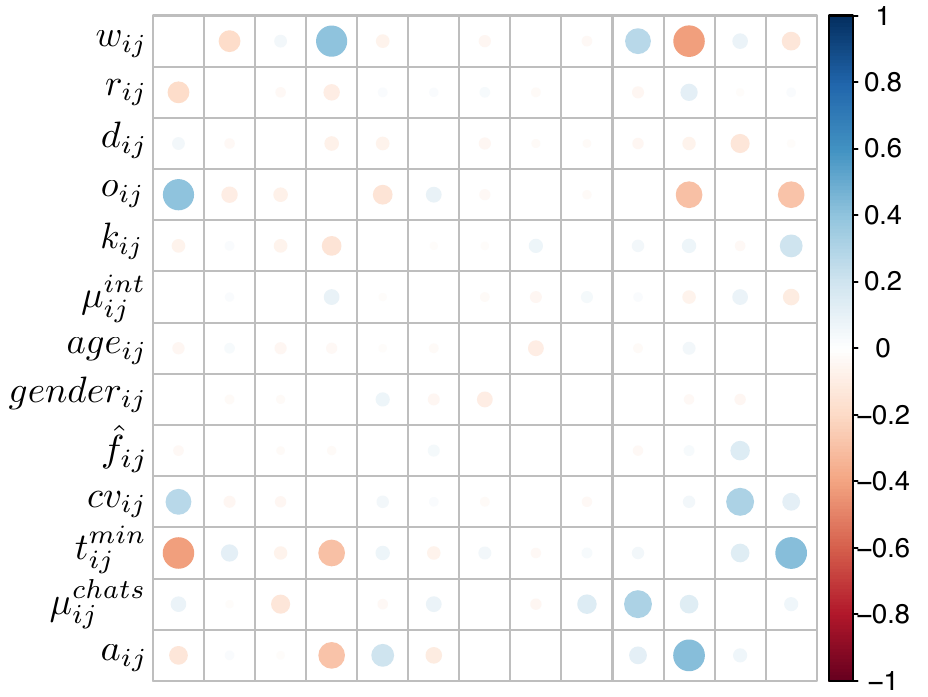}
\caption{{\bf Correlation between features.} Correlation matrix for the different tie features considered in the model. Each entry shows the Pearson correlation coefficient between two variables. Size is proportional to the absolute value of the correlation coefficient, while color shows also its sign. We only show correlation coefficients which are significantly different from zero (with a 95\% confidence interval).}
\label{fig:corrmat}
\end{figure*}

Finally, the variables constructed might be all relevant to our predicting model, but they can carry redundant information about the ties, i.e., they can be highly correlated. It is well known that correlated variables can diminish the predicting power of the model and thus we must understand the explanatory power between them first in order to construct a statistical significant model. This process which is known as selection of variables will be address qualitatively in this section using the correlation matrix between them. As we can see in figure \ref{fig:corrmat} most of the variables we have selected are highly uncorrelated. The only exception being the relationship between number of calls and topological overlap, i.e. the Granovetter effect \cite{Granovetter1973,Onnela:2007p316}. Since correlations between the rest are moderate, we keep all features in our analysis.

\subsection{Facebook data}
We have also analyzed other communication data to test the independence of our results to the particular mobile phone setting. In particular, we have studied the 90,269 users of the New Orleans Network crawled during December 29th, 2008 and January 3rd, 2009 by Vismanath {\em et al.} \cite{Viswanath:2009:EUI:1592665.1592675}. The data consists of communication events between users through Facebook wall. Contrary to the mobile phone data, the Facebook data is not steady in time, since the database extends over the early days of Facebook growth and thus it shows a growth in the activity over years, which translates in more wall posts and also more users as a function of time.  

To minimize this effect we have chosen only communication events between users that did show any activity in the observation window $\Omega$ (the time interval between 1000 and 1212 days in the database) and also which were present 20 days before and after $\Omega$. We do not consider the ties to be reciprocated in order to have more data accessible for our analysis. With this filter our database contains $125 \times 10^3$ communication events of $\sim 10^4$ users and $69 \times 10^3$ ties. We have considered only 5466 ties which are more active (more than 5 communication events) and build a predictive model similar to the one for the mobile phone data. However, since we do not have information about the age and gender of the users, we have discarded the variables related to their difference. Results of our model for the Facebook data are presented in table \ref{table:glmmodelsFB} where we can see that qualitatively that they match the ones for the mobile dataset, although the predictive power of the models is smaller than in that case. Apart from the number of communication events, both the normalized freshness and the coefficient of variation have a similar relevant role in predicting tie persistence. In particular, we find that the critical relative freshness is now $\hat f_{ij} = 16.6$, which is double that the one found in the mobile phone calls. This could be a signature of the different rhythm of communication of users on different channels.

\begin{table*}[!ht]
\centering
\caption{Regression results for the tie persistence using generalized linear models for the two prediction models in the Facebook dataset. Coefficients are shown with uncertainties (standard errors) in parentheses. Model {\em Full} include all the features described in the text, while model {\em Simplified} only includes the most important two features.}
\begin{tabular}{ c | c  c | c c | }
				& 		\multicolumn{2}{c}{\textbf{Model 1}} & \multicolumn{2}{c}{\textbf{Model 2}} \\ 
\textbf{Feature} & Full & Simplified  & Full & Simplified  \\ \hline
\hline
$w_{ij}$ 		& $0.839^{***}$	& $1.228^{***}$ & $1.709^{***}$	& $2.041^{***}$		\\ 
				& (0.044) 		& (0.066)		& (0.180)		& (0.105)	 \\
$r_{ij}$ 		& $0.118^{***}$&				& $0.470*{***}$	&				\\
				& (0.034) 		& 				& (0.111)		& 			\\
\hline
$o_{ij}$ 		& $0.269^{***}$	& 				& $0.384^{***}$	& 			\\
				& (0.035) 		&				& (0.111)		& 				\\
$k_{ij}$ 		& $-0.008$		&  				& $-0.231^{*}$	& 				\\
				& (0.031) 		&				& (0.099)		& 			\\
\hline
$\mu_{ij}^{int}$ & $-0.0139^{}$	&  				& $0.006$		& 		\\
				& (0.030) 		& 				& (0.095)		& 		 \\
\hline
$\hat f_{ij}$ 	& $-0.608^{***}$& $-0.306^{***}$&				& 	 \\
				& (0.0390) 		& (0.016)		&				& 	\\
$t_{ij}^{min}$ 	& $-0.329^{***}$&  				& $-0.402^{**}$& 	\\
				& (0.039) 		& 				& (0.128)		& 	\\
$cv_{ij}$ 		& $-0.286^{***}$&  				& $-0.229^{*}$&  $-2.394^{***}$\\
				& (0.037) 		& 				& (0.103)		&  (0.224) \\
$\mu_{ij}^{chats}$ & $0.024^{}$& 				& $-0.047^{}$	& 			\\
				& (0.035) 		& 				& (0.114)		& 				\\
$a_{ij}$ 		& $0.122^{***}$&  				& $0.115^{}$	& 				\\ 
				& (0.036) 		& 				& (0.111)		& 				\\ \hline
Constant 		& $0.435^{***}$ & $1.951^{***}$	& $0.686^{***}$	& $-4.525^{***}$ \\
				& (0.032) 		& (0.155)		& (0.115)		& (0.265) \\
\hline 
Number of observations & 5466		& 5466		& 667			& 667 \\
\hline
\textbf{Performance} & 			& 				& 				& 				 \\ \hline
Accuracy 		& 0.690 		& 0.688			&  0.798		& 0.799				 \\
Sensitivity 	& 0.770 		& 0.780			&  0.814		& 0.802				\\ 
Specificity 	& 0.583 		& 0.567	 		&  0.776		& 0.797					\\
\hline
\multicolumn{5}{l}{\textit{Note:} $^{*}p<0.1$;  $^{**}p<0.05$; $^{***}p<0.01$}  \\
\end{tabular}
\label{table:glmmodelsFB}
\end{table*}

\section*{Competing interests}
  The authors declare that they have no competing interests.

\section*{Author's contributions}
  All authors contributed equally to this work.

\section*{Acknowledgements}
We would like to thank Telef\'onica for providing access to the anonymized data. E.M. acknowledges funding from Ministerio de Econom\'{\i}a y Competividad (Spain) through projects FIS2013-47532-C3-3-P and FIS2016-78904-C3-3-P.
  

\bibliographystyle{apalike}

\end{document}